# M-I Transition in a-Conducting Carbon Films Induced by Boron Doping

*P.N.Vishwakarma\* and S.V.Subramanyam*
*Department of Physics*
*Indian Institute of Science*
*Bangalore -560012 (India)*

Abstract:

The amorphous conducting carbon films have been prepared at three different preparation temperatures with different boron-doping levels. The structural and transport properties of the same have been studied. X-ray diffraction measurements show that the 'd' value of the carbon depends both on atomic percentage of B in the carbon network and also on the preparation temperature. Doping of boron increases the structural graphitic ordering of the films prepared at lower temperatures. On the contrary for the films prepared at higher temperatures the ordering deteriorates as the boron content increases. The d.c electrical transport measurements on these amorphous conducting carbon films show, doping induced metal-insulator transition via critical regime, in the temperature interval of 1.3 K to 300K. Also the films in the insulating regime show a crossover from Mott to ES VRH for T < 55K. Additional support to this transition is evident from negative magnetoresistance in VRH regime when the sample is deep inside the insulating side of MI transition. The calculated value of density of states at Fermi level shows a gradual change with corresponding variation in boron doping level, indicating a change in the number of conducting pi electrons due to substitutional doping of boron in the carbon network. However for the films exhibiting critical behaviour, the magnetic field dependence of magnetoresistance was not as predicted by the available theoretical models. Various calculated parameters like localization length, density of states at the Fermi level and coulomb gap for insulating samples were calculated from the experimental data.

*\*email: prakash@physics.iisc.ernet.in*

## I. INTRODUCTION

The density of states (DOS) for disordered metals show localized behaviour near the band edge [1]. Since the electrons taking part in conduction process are near the Fermi level $E_F$, a disordered metal can be tuned to insulating state if the states near the Fermi level are localized. This can be done either by increasing disorder in the system so that states upto Fermi level are localized or by changing the carrier density so that the Fermi level falls below the band edge. Recent studies on amorphous conducting carbon films (a-cc) have shown disorder induced M-I transition by varying the preparation temperature of the sample [2]. Also a crossover from Mott VRH to Efros-Shklovskii VRH conduction at low temperatures was observed for insulating samples (prepared at $700^0$C). Such crossover from Mott VRH to Efros-Shklovskii VRH conduction has also been observed in mesoscopic carbon networks and the crossover temperature is very much nearer to that observed in our system [3]. The presence of coulomb gap as predicted by Efros-Shklovskii is also seen in transport measurements done on granular and porous carbon structures [4]. Boron atom has one electron less than the carbon atom and C-B bond is longer than C=C bond. So the effect of boron doping in the carbon network will be to bring structural changes due to mismatch in bond lengths and decrease in the density of conducting $\pi$ electrons thereby bringing down the Fermi level. In the early work of boron doping in carbon, it was found that boron doping induces graphitization in the carbon network [5].

Here we are reporting doping induced MI transition in amorphous conducting carbon films using boron as a dopant.

## III. EXPERIMENTAL DETAILS

The Boron doped amorphous conducting carbon films were prepared using high temperature pyrolysis assisted CVD method. Standard solution of boric acid ($H_3BO_3$) in distilled water was used as the boron source [6]. One ml of 1Molar boric acid solution along with 300mg of maleic anhydride ($C_4H_2O_3$) was taken in a 50cm x 1cm quartz tube closed at one end. The other end of the tube was closed with a rubber bladder. Boron concentration in the sample was varied by varying the initial concentration of boric acid solution (1M, M/2, M/4). For the notation of sample CB900M2, C and B stands for carbon and boron, 900 is the pyrolysis temperature in $^0$C and M2 stands for half molar boric acid solution initially taken. The films were deposited over fused quartz substrates (5mm x 5mm), cleaned in water with few drops of HF and in acetone prior to deposition. The quartz tube was heated until the center of the furnace reached the pyrolysis temperature ($900^0$C / $800^0$C / $700^0$C) and then the precursors were allowed through the hot zone for 5 hours and then the furnace was allowed to cool in normal ambient conditions. The quartz tube was pre-evacuated and purged with inert gas prior to pyrolysis. Adhesion of the films deposited on the inner wall of the tube was very good. So to take out the boron doped films, the quartz tube was filled with water with few drops of HF and left for few hours until the films came out of the inner wall. These films were taken out and subsequently dried and made into powder for XRD analysis. The boron doped carbon film deposited on the substrate was taken out for electrical transport measurements. The thickness of the film deposited over the quartz substrate was approximately 1 $\mu m$. Resistivity measurements were done in the temperature range of 300K-1.3K and in the magnetic field up to 7 tesla using Janis Supervaritemp Superconducting Magnet Cryostat. Electrical contacts to the samples were done using conducting silver epoxy.

## IV. RESULTS AND DISCUSSION

Substitution of a boron atom into carbon layer leads to formation of three carbon-boron (C-B) bonds and that the bond length of C-B (0.148 nm) is a little longer than that of C=C (0.142 nm), so the average bond length in the layer plane will be longer than that of C=C. The other effect of boron substitution in the carbon network will be to decrease the π electron density. The simple van der waals force between the two layers in graphite lattice gives an interlayer spacing of 0.344 nm and the interaction between the π-electron clouds of the two layers reduces the interlayer spacing to 0.3354 nm. Due to boron doping the π electron density decreases which gives rise to increase in interlayer spacing due to reduction in electron interaction. One more factor, which contributes to the change in interlayer spacing, is Poisson contraction. This effect predicts decrease in interlayer spacing due to inplane stretching when boron is substituted in the carbon network. So the net effect on the interlayer spacing will be combination of both the factors. The net change in lattice parameters due to boron doping in a graphite lattice can be written as [7],

$$\Delta a_0 = a - a_0 = 0.000310 x_B \quad \text{and} \quad \Delta c_0 = c - c_0 = -0.000594 x_B \quad (1)$$

where $\Delta a_0$ and $\Delta c_0$ (in nm) are the changes in lattice constants with respect to the atomic fraction of boron $x_B$ (atomic percentage of doped boron) in the carbon network. Though our carbon is not as ordered as graphite, we have used the above relation to find out approximate boron concentration in the sample.

Fig.1 shows the X-ray diffraction pattern of boron-doped amorphous conducting carbon prepared at different temperatures and for various boron concentrations. XRD patterns of all the samples exhibit mainly two types of peaks: (002) resulting from stacks of parallel layer planes at 23-29$^o$ and two dimensional (10) peaks resulting from the regular structure within the individual layer plane segments at 41-47$^o$. Peaks of the type (hkl) are absent, indicating that there is little or no stacking order in the arrangement of parallel layers. These patterns are known to be characteristic of disordered graphite like structure of carbon where the graphene layers are regularly stacked in itself but has no correlation to the next pile except for parallelism[8, 9]. For the samples prepared at 800$^0$C, boron-doping increases the intensity of the peak at 25$^0$ with subsequent decrease in FWHM of the same, which is an indication of graphitic ordering in the carbon system. XRD spectra for samples prepared at 700$^0$C are similar to those of 800$^0$C, so they are not shown in the figure. Such doping induced graphitization has been observed in other forms of disordered carbon systems[5, 10]. However, for the carbon films prepared at 900$^0$C, boron doping introduces disorder in the carbon system without bringing any graphitization, which is reflected in the decrease in intensity of peak at 24$^0$ and broadening of FWHM for the same (table I). Such boron-induced disorder has been observed in few carbon systems where the boron induces graphitization but only upto certain critical value of dissolved boron in the carbon network. Any further introduction of boron in the carbon lattice brings disorder in the carbon lattice instead of graphitization. So if a carbon system is in its maximum structural ordering then the boron doping in that carbon network will deteriorate the structural ordering instead of bringing out any graphite ordering. Such boron doping induced structural disordering has also been observed in HOPG carbon systems [7]. A relatively intense peak and less FWHM for the same in x-ray diffraction data of C900 and C800 suggests that C800 films are more graphitic disordered than C900. Probably the carbon films C900 are at maximum graphitic ordering and cannot be graphitized any further using external dopant. Hence boron substitution in these carbon films introduces disorder instead of graphitic ordering. Table I shows a comparison between the various carbon samples, analyzed using XRD. Maximum boron concentration for the samples prepared at $800^0$C is 25% (table I). The atomic boron concentration for the samples prepared at $900^0$C and $700^0$C is not tabulated and assumed to be approximately the same as the initial concentration of boric acid solution was same for all the three sets of samples (i.e., $700^0$C, $800^0$C and $900^0$C).

Following the common practice adopted in literature, we will discuss the data for metallic samples in terms of conductivity ' $\sigma$ ' and magnetoconductance $\Delta\sigma = \sigma(H) - \sigma(0)$ and for critical and insulating samples in terms of resistivity ' $\rho$ ' and relative magnetoresistance $\rho(H)/\rho(0)$ .

Classifications of metallic and insulating samples were done using reduced activation energy 'W' ($W(T) = -\partial(\log\rho)/\partial(\log T)$ plots as suggested by Zabrodskii [11]. The reduced activation energy plots for the samples prepared at $900^0$C and $800^0$C are shown in figure 2. To obtain these plots, we have smoothened the $\ln\sigma$ vs lnT data and then differentiated the data to get 'W'. This was done to avoid scattering due to differentiation. The 'W' plot for samples prepared at $700^0$C are similar to those of CB800M samples, and are not shown here.

According to the classification scheme discussed by Zabrodskii, C900 sample shows metallic behavior whereas CB900M, CB900M2, CB800M, C700 and CB700M4 show insulating behaviour. For samples C800, CB900M4, CB800M2 and CB800M4 the reduced activation energy curve shows metallic behaviour at higher temperatures but at lower temperatures it shows tendency towards insulating behaviour. In these samples the Fermi level lies very close to the mobility edge and at higher temperatures the 'thermal smearing' of Fermi level (i.e., spread of the order of $k_B T$ in the Fermi energy due to thermal vibrations) effectively shifts it away from the mobility edge, thereby suppressing the critical character of the material.

## A.  Metallic Regime

The temperature dependence of conductivity for C900 (metallic) sample is shown in Fig 3. The electrical conductivity in the disordered metals is very well explained by the theory of weak localization [12] with interaction correction at low temperatures,

$$\sigma = \sigma_0 + \frac{e^2}{\hbar}\frac{1}{\pi^2}\frac{T^{p/2}}{a} + \frac{e^2}{4\pi^2\hbar}\frac{1.3}{\sqrt{2}}\left(\frac{4}{3} - \frac{3}{2}\tilde{F}_\sigma\right)\frac{\sqrt{T}}{D} \qquad (2)$$

where $a$ is some microscopic length scale such as $k_F^{-1}$ and D is the Einstein diffusion coefficient ( $D = e^2 N(E_F)/\sigma_0$ ). The first term $\sigma_0$ is Drude conductivity, second term comes due to weak localization effect and the last term arises due to electron-electron interaction effect. The value of $p$ determines the strength of coulomb interaction ( p is 3/2 when interaction is in the dirty limit, 2 when it is in clean limit and 3 when electron-phonon scattering  determines the inelastic scattering rate) [13].

So at low temperature (T < 30K) we have fitted the following equation[7] to the conductivity data.

$$\sigma(T) = \sigma_0 + mT^{1/2} + BT^{p/2} \qquad (3)$$

where the first term on the right hand side of equality is Drude conductivity, the second term comes due to electron-electron interaction and the third term is due to weak localization. The values of different parameters are. $\sigma_0$ = 390 S/cm, m=10, B=0.12 and p =2.93. The value of p $\approx$ 3 indicates that the electron-phonon scattering is the main dephasing mechanism. Below 10 K the corresponding values are $\sigma_0$ = 398 S/cm, m=3.18, B=1.38 and p =2. So below 10 K electron-electron interaction (p =2) is playing an important role. At high temperatures the conductivity varies as $T^{1/2}$. Such temperature dependence was also observed in metallic glasses by Howson [14] for T > $\theta_D$ /3. The reported values of $\theta_D$ for amorphous carbons, which have structures and density similar to our carbon samples, are around 310K [9]. Therefore we believe the $T^{1/2}$ variation of conductivity at high temperatures is due to weak localization effect.

The presence of magnetic field suppresses the weak localization effect, which subsequently enhances the conductivity and one can observe negative magnetoresistance (or positive magnetoconductance). The magnetoconductance plot in Fig 4 is showing $H^2$ and $H^{1/2}$ dependence at low and high magnetic field respectively as predicted by theory [15]

$$\sigma(B) - \sigma(0) = \frac{e^2}{12\pi^2\hbar}\frac{1}{a}T^{p/2}\left(\frac{eB}{\hbar}\right)^2 - \frac{e^2}{4\pi^2\hbar}\tilde{F}\left(\frac{0.056g\mu B}{2\hbar D}\right)^2 \qquad ;(g\mu_B B << k_B T) \quad (4)$$

$$\sigma(B) - \sigma(0) = \frac{0.605e^2}{2\pi^2\hbar}\left(\frac{eB}{\hbar}\right)^{1/2} - \frac{e^2}{4\pi^2\hbar}\tilde{F}\left(\frac{g\mu B}{2\hbar D}\right)^{1/2} \qquad (g\mu_B B >> k_B T) \quad (5)$$

where $\mu_B$ is Bohr magneton. The first term (in both the equations) is due to weak localization and the second term is due to electron-electron interaction. As the temperature falls, thereby reducing $\tau_i$, the $B^{1/2}$ dependence persists to lower and lower fields and at absolute zero of temperature, the magnetoconductance varies as $B^{1/2}$ over the full range of field. There is a flip in the direction of magnetoconductance from positive to negative at temperatures below 10 K. The positive magnetoconductance at 10K is characteristic of weak localization whereas it is negative for lower temperatures when electron-electron interaction is playing a dominant role.

## B. Insulating Regime

Assuming that density of states (DOS) near the Fermi level $N(E_F)$ is a slow varying function of energy, Mott [12] first predicted the hopping form of electrical resistivity for a three-dimensional (3D) insulating material as

$$\rho(T) = \rho_{Mott} \exp\left(\frac{T_{Mott}}{T}\right)^{\frac{1}{4}} \quad (6)$$

where the characterstic temperature $T_{Mott}$ is found from the slope of the $\ln \rho$ versus $1/T^{1/4}$ data. $T_{Mott}$ is related to the desity of states at Fermi level $N(E_F)$ and to the localization length $(\alpha^{-1})$ as

$$T_{Mott} = \frac{18\alpha^3}{k_B N(E_F)} \quad (7)$$

For the Mott theory to be valid, the electron must hop a mean distance $R_{hop,Mott}$ that is considerably greater than the nearest neighbour impurity separation and considerably greater than the localization length $(\alpha^{-1})$. The mean hopping distance $R_{hop,Mott}$ and the mean hopping energy difference $\Delta_{hop,Mott}$ between sites can be written as [16, 17],

$$R_{hop,Mott} = \frac{3}{8}\alpha^{-1}(T_{mott}/T)^{1/4} \text{ and } \Delta_{hop,Mott} = \frac{1}{4}k_B T(T_{Mott}/T)^{1/4} \quad (8)$$

While deriving the expression for conductivity in insulating regime, Mott did not consider the interaction between hopping sites. However this interaction effect was later discussed by Pollak, Srinivasan, Efros Shklovskii and Shklovskii and Efros [18-21]. This interaction effect predicts power law dependence in the density of states so that DOS vanishes at Fermi level thereby creating a gap at $E_F$.

$$N(E) = N_0 \mid E - E_F \mid^\gamma \tag{9}$$

where the exponent $\gamma = 2$ for 3D films. The expression for this Coulomb gap $\Delta_{CG}$ is

$$\Delta_{CG} = k_B T_{CG} \tag{10}$$

Efros-Shkolovskii predicted the following temperature dependence of resistivity due to interaction effects for all dimensions [18, 19].

$$\rho(T) = \rho_{ES} \exp\left(\frac{T_{ES}}{T}\right)^{\frac{1}{2}} \tag{11}$$

The corresponding mean hopping distance $R_{hop,ES}$ and the mean hopping energy difference $\Delta_{hop,ES}$ between sites for Efros-Shkolovskii VRH conduction is

$$R_{hop,ES} = \frac{1}{4}\alpha^{-1}\left(T_{ES}/T\right)^{1/2} \text{ and } \Delta_{hop,ES} = \frac{1}{2}k_B T(T_{ES}/T)^{1/2} \tag{12}$$

Castner predicted the following relations for a material undergoing transition from Mott to ES VRH conduction [17, 22],

$$T_{Mott}/T_{ES} = 18(4\pi)/\beta_1 \approx 81 \quad \text{for } \beta_1 = 2.8 \tag{13}$$

$$T_{Mott}/T_{CG} = 18(4\pi)^{3/2} \approx 801, \tag{14}$$

$$T_{ES}/T_{CG} = \beta_1(4\pi)^{1/2} \approx 9.9 \quad \text{for } \beta_1 = 2.8 \tag{15}$$

For Samples in insulating regime, the R vs T data were analyzed using a procedure suggested by Zabrodskii and Zinov'era [23]. In this regime the general expression for resistivity in VRH conduction can be written as $\rho(T) = \rho_0 \exp(T_0/T)^x$, where x = 0.25 for Mott VRH conduction and x = 0.5 for Efros-Skhlovskii coulomb interaction. Then reduced activation energy (W) which is defined as $W(T) = -\partial(\log\rho)/\partial(\log T)$ can be written as $W(T) = x(T_0/T)^x$ and $\log W(T) = \log(xT_0^x) - x\log T$, so that it can be fitted to a linear equation $Y = C - mX$ where, $Y = logW(T)$, $X = logT$ and $C = \log(xT_0^x)$.

For samples CB900M, CB900M2, CB800M, CB700M4 and C700 at low temperatures the exponent x = 0.5 gives a good fit whereas for T > 20K (T > 50K for CB700M4 and C700), x = 0.25 gives a better fit (Fig 2) indicating a crossover from Efros-Shklovskii (ES) VRH to Mott VRH for T > 20K. The intersection of these two straight lines gives the crossover temperature $T_{cross}$. This crossover occurs when $\Delta_{hop,Mott} \approx \Delta_{hop,ES}$. The expression for $T_{cross}$ derived using Eq.(8) and Eq.(12) is, $T_{cross} = 16T^2_{ES}/T_{Mott}$.

The temperature dependence of resistivity for CB900M, CB900M2, C700, CB700M and CB800M is plotted in Figure 5. Similar to reduced activation energy plot, resistivity is giving a good fit for x = 0.5 at low temperature and x = 0.25 at high temperature. The $T_{cross}$ is determined by the intersection of the linear fit for x = 0.25 and x = 0.5. The $T_{Mott}$ value for all the samples were calculated from resistivity data as it gives a longer x-axis range. The value of $T_{Mott}/T_{ES}$ is very low as compared to that predicted by Castner [Eq.(13)]. Such a low value has also been observed for system showing $T_{cross}$ as low as ours [24]. Low value of $T_{Mott}/T_{ES}$ had been attributed to the inability to determine $T_{Mott}$ with much accuracy. The theoretically predicted value of the ratio $T_{Mott}/T_{ES}$ is derived for highly insulating materials ( $\beta_1 = 2.8$ ), whereas the experimental reported values of $T_{Mott}/T_{ES}$ are for relatively conducting samples. So we attribute the low value of $T_{Mott}/T_{ES}$ to the conductivity of the films as compared to those considered in theoretical modeling.

Once $T_{Mott}$ is known, the coulomb gap can be calculated using Eq. (10) and Eq. (14). Through there are reports of calculating coulomb gap using Eq. (15), we prefer Eq. (14) for calculating the coulomb gap because it does not contain the term $\beta_1$, which is unknown for our samples. The values of coulomb gap for the samples showing ES VRH conduction mechanism, are tabulated in table II, III and IV. The coulomb gap for samples C700 and CB700M4 are well above the thermal energy kT (0.36meV), whereas that for CB900M, CB900M2 and CB800M is either comparable or lower than kT. So for samples C700 and CB700M4, the transition in conduction mechanism can be attributed to the coulomb gap. Although the samples CB900M, CB900M2 and CB800M are showing ES behaviour at low temperatures, it is difficult to admit the existence of gap due to such a low value of coulomb gap. However such comparable value ($\Delta_{CG}$ near to kT) of coulomb gap has also been observed using transport data [24]. Although films that exhibit the crossover from Mott to ES VRH conductivity do suggest the existence of the coulomb gap, these transport measurement in no way provide direct evidence for the existence of the gap. Recently few groups were able to detect this coulomb gap using tunneling and spectroscopic measurements on various samples [25, 26].

The effect of magnetic field in a VRH system is to induce shrinkage of the wavefunction thereby reduction in overlapping of wavefunction. Hence there is a decrease in the probability of tunneling/hopping and increase in the resistance. The expressions for magnetoresistance in the two limiting cases are [18]

$$\ln\left(\frac{\rho(H)}{\rho(0)}\right) = 0.00248 \frac{e^2 H^2}{c^2 \hbar^2 \alpha^4}\left(\frac{T_{mott}}{T}\right)^{3/4} \quad (\lambda < \alpha^{-1})$$ (16)

$$\ln\left(\frac{\rho(H)}{\rho(0)}\right) = \exp\left\{\left(2.1\frac{e\alpha_H H}{N(E_F)c\hbar k_B}\right)^{1/3}\left(\frac{1}{T}\right)^{1/3}\right\} \quad (\lambda > \alpha^{-1})$$ (17)

where $\lambda = (c\hbar / eH)^{1/2}$ is magnetic length.

Similarly the expression for magnetoresistance in insulating regime due to interaction effects in the two limiting cases are,

$$\ln\left(\frac{\rho(H)}{\rho(0)}\right) = 0.0015 \frac{e^2 H^2}{c^2 \hbar^2 \alpha^4}\left(\frac{T_{ES}}{T}\right)^{3/2} \qquad\qquad ;(\lambda < \alpha^{-1})$$ (18)

$$\ln\left(\frac{\rho(H)}{\rho(0)}\right) = \exp\left\{s\left(\frac{eH}{c\hbar\alpha_H^2}\right)\left(\frac{T_0}{T}\right)^{3/5}\right\}, \text{where} \quad T_0 = \frac{3.17e^2}{\kappa}\left(\frac{\alpha_H}{\lambda^2}\right)^{1/3};(\lambda > \alpha^{-1})$$ (19)

In general Eq.(16) and Eq.(18) can be written as

$$\ln\left(\frac{\rho(H)}{\rho(0)}\right) = AH^2$$ (20)

Taking log on both sides,

$$\log\left[\ln\left(\frac{\rho(H)}{\rho(0)}\right)\right] = \log A + 2\log H$$ (21)

Figure 6 shows the plot of log of magnetoresistance as square of magnetic field at low magnetic fields and linear dependence of magnetic field at high fields for samples showing crossover from Mott to Efros-Shklovskii VRH as predicted in Eq.(18) and Eq.(19). The localization length ($\alpha^{-1}$) was calculated by measuring the intercept for the graph log[lnR(H)/lnR(0)] vs logH and using equation Eq.(21). For the samples showing crossover, the value of 'A' was taken from Eq.(18). The respective values of localization length ($\alpha^{-1}$) for samples showing insulating behaviour are tabulated in table II, III and IV.

The gradual decrease in the values of $\alpha^{-1}$ with increasing resistivity is the signature of localization. The mean hopping distance $R_{hop,Mott}$ and $R_{hop,ES}$ derived from $\alpha^{-1}$ is also tabulated in table II, III and IV.

These values of $\alpha^{-1}$ along with $T_{Mott}$ value were further used to calculate $N(E_F)$

at 4.2K using Eq. (5) for films showing Mott behaviour. The value of density of states is coming down, as the sample is becoming more insulating. This is expected as increasing content of boron in the carbon lattice reduces the conducting $\pi$ electrons.

For samples C700, CB700M4 and CB700M2, lying deep inside the insulating

side of MI transition, the magnetoresistance behaviour is shown in figure 6. For sample

CB700M2, there is a transition in magnetoresistance at 2 tesla and 20K. The

magnetoresistance first goes to negative values and then at 2T again it goes towards

positive values, whereas at 15K the magnetoresistance is negative throughout the range

of applied magnetic field. Such phenomenon has also been observed in doped

semiconductors and copper indium diselenide [27]. The theoretical aspect of negative

magnetoresistance in hopping conduction has been analyzed by Nguyen's et.al [28] and

Altshuler et.al.[29]. In both the models, the enhanced electron-electron interaction has resulted in negative magnetoresistance in hopping conduction. The negative magnetoresistance observed in hopping conduction is a clear indication of formation of coulomb gap at that temperature.

<div style="text-align:center">C. Critical Regime:</div>

In the critical state, the localization length $\alpha^{-1}$ is large enough so that over distances $r << \alpha^{-1}$ the electrons behave as free particles. For this critical regime Macmillan [30] and Larkin and Khmel'nitskii [31] gave an expression for the conductivity in critical regime,

$$\rho(T) = \left(\frac{\hbar^2}{e^2 p_F}\right)\left(\frac{E_F}{k_B T}\right)^{\beta} \Rightarrow \rho(T) \propto T^{-\beta} \qquad (22)$$

where $0.33 < \beta < 1$ and $p_F$ is the Fermi momentum. The above relation has been derived taking into account electron-electron interactions. If these interactions are absent, $\beta = 1$. The samples C800, CB800M4, CB800M2 and CB900M4 were analyzed for the critical behaviour using Eq. (22). For the sample CB900M4 and CB800M2, which shows critical behaviour, the value of $\beta$ is very close to 0.25 (Figure 5). This is contrary to the theoretical prediction, *0.33< β < 1*. Such a low value of $\beta \approx 0.25$ has also been observed in n-Ge [32] and doped polyaniline [33-35]. Such a low value of $\beta$ is considered to be an indication that sample is very close to metal-insulator boundary, but on the metallic side. For CB800M4, $\beta$ =0.6 as expected according to the theory. C800 gives $\beta$ =0.95, indicating absence of interaction effects as discussed previously.

   At present there is no microscopic theory for the effect of magnetic field on the systems in the critical regime. Based on scaling argument, Khmelnitskii and Larkin [36] gave an expression for the magnetoconductance for critical samples where $\Delta\sigma \propto H^{1/2}$. Figure 7 shows the plot of magnetoconductance vs $H^{1/2}$. From Figure 7 it is evident that the magnetoconductance data is not behaving as per prediction by Khmelnitskii and Larkin. Probably the magnetic field applied here is too low as for the theoretical prediction. We found that in this magnetic field value the magnetoresistance goes as $\ln(R(H)/R(0)) \propto H^{3/2}$ (Figure 7). As per our information at present there is no theoretical model available, which predicts such dependence of magnetoresistance on the magnetic field.

## V. CONCLUSIONS

Boron doped amorphous conducting carbon films were successfully prepared without segregation of boron in carbon films even for 25 at% of boron in the amorphous conducting carbon film. Boron doping in the amorphous conducting carbon films prepared at $800^0$C and $700^0$C, bring graphitic ordering whereas for the amorphous conducting carbon films prepared at $900^0$C structural disordering was observed due to boron doping. This is probably due to maximum ordering state achieved by C900 and it cannot be graphitized further using external dopant. As discussed earlier, the other effect of boron in the carbon network is to decrease the density of conducting $\pi$ electrons. This is seen as a gradual change in $N(E_F)$ with increasing concentration of boron in the amorphous conducting carbon film (Table II, III and IV). Since electrical conductivity in a disordered system depends both on the extent of disorder as well as the density of conduction electrons, the net effect of boron doping in the carbon films prepared at $800^0$C is first to decrease the temperature dependence of resistivity due to graphitization (ordering) in the system and then it drives the system towards insulating state due to low density of conducting $\pi$ electrons with increasing boron concentration in amorphous conducting carbon film. However in the samples prepared at $900^0$C, graphitic ordering due to boron doping is not happening due to metastable ordered structure already acquired, so the effect of boron doping in the amorphous conducting carbon prepared at $900^0$C is to drive the system towards insulating state by increasing disorder and decreasing the density of conducting $\pi$ electrons. The values of various calculated parameters like localization length, coulomb gap, crossover temperature, etc for B doped amorphous conducting carbon are tabulated in tables II, III and IV. So based on the parameters calculated from experimental data, it is evident that MI transition has been observed in B doped amorphous conducting carbon. Boron doping is driving the amorphous conducting carbon system from metallic to deep insulating state, so that coulomb gap is also observed. Among the films showing insulating behaviour, coulomb gap for CB900M, CB900M2 and CB800M was found to be less than the kT but still ES behaviour is seen in the conductivity data. This discrepancy is not very clear. For the samples C700 and CB700M4, coulomb gap is well above kT and negative magnetoresistance which results from enhanced electro-electron interaction, is also observed. Since transport measurement is not a foolproof technique to quantify the energy gap, optical measurements on these samples can give some more information about coulomb gap. Also for the critical samples the magnetoresistance is not varying as predicted theoretically rather it is varying as $H^{3/2}$. May be the applied magnetic field is low for the theoretical prediction and will be observed at higher magnetic fields.

# REFERENCES

1. *Electronic process in non-crystalline materials*, N.F.Mott and E.A.Davis, (Clarendon press, Oxford, 1971), p200; N.F.Mott, Phil. Mag. **19**, 835(1969)

2. Ahmed Sayeed, *Ph.D. thesis*, Indian Institute of Science, (1998)

3. V.A. Samuilov, J. Galibert, V.K.Ksenevich, V.J.Goldman, M.Rafailovich, J.Sokolov, I.A.Bashmakov, V.A.Dorosinets. *Physica B*, **294-295**, 319(2001).

4. A.W.P.Fung, Z.H.Wang, M.S.Dresselhaus, G.Dresselhaus, R.W.Pekala, M.Endo. *Phys. Rev. B*. **49**, 17325(1994)

5. S.Marinkovic in *Chemistry and Physics of Carbon*, ed. by P.A.Thrower, vol.21,p1-64, (Marcel Dekker: New York, 1988)

6. J. T. Huang, W. H. Guo, J. Hwang, and H. Chang, *Appl. Phys. Letts*. **68**, 3784(1996)

7. Y.Hishiyama, H.Irumano and Y.Kaburagi, *Phys. Rev. B* **63**, 245406(2001)

8. R.E.Franklin, *Acta. Cryst* **3**,107(1950)

9. W.Ruland in *Chemistry and Physics of Carbon*, ed by P.L.Walker Jr, vol.4,p1-84, (Marcel-Dekker: New York, 1968)

10. W. Cermignani, T.E.Paulson, C.Onneby and C.G.Pantano, *Carbon* **33**, 367(1995).

11. A.G.Zabrodskii, *Sov. Phys. Semicond*. **11**, 345(1977)

12. E.Abrahams, P.W.Anderson, D.C.Licciardello and T.V.Ramakrishnan, *Phys. Rev. Lett*. **42**, 673(1979)

13. P.A.Lee and T.V.Ramakrishnan, *Rev. Mod. Phys* **57**, 287(1985).

14. M.A.Howson, *J. Phys F: Met. Phys*. **14**, L25(1984)

15. *The Electrical Properties of Disordered Metals*, J.S.Dugdale, (Cambridge University Press, Cambridge, 1995).

16. N.F.Mott, *J.Non-Cryst. Solids* **1**,1,(1968)

17. T.G.Castner, *Hopping Transport in Solids*, edited by M. Pollak and B. I.

    Shklovskii (Elsevier/North-Holland, Amsterdam, 1990), p.1.

18. B. I. Shklovskii and A. L. Efros, in *Electronic Properties of Doped Semiconductors*,

    Edited by M. Cardona (Springer-Verlag, Berlin, 1984)

19. A. L. Efros and B.I.Shklovskii, *J. Phys. C* **8**, L49 (1975).

20. M. Pollak, *Discuss. Faraday* Soc. **50**, 13(1970).

21. G. Srinivasan, *Phys. Rev. B* **4**, 2581 (1971)

22. W.N.Shafarman, D.W.Koon and T.G.Castner, *Phys. Rev. B* **40**, 1216(1989)

23. A.G.Zabrodskii and K.N.Zeninova, *Sov. Phys. JETP* **59**, 425(1984)

24. R. Rosenbaum, *Phys. Rev. B* **44**, 3599(1991)

25. J.G.Massey and Mark Lee, *Phys. Rev. B* **62**, R13270 (2000)

26. R.C.Dynes and J.P. Garno, *Phys. Rev. Lett* **46**,137(1981); W.L.MacMillan and

    J.Mochel, *Phys. Rev. Lett* 46,556(1981)

27. L. Essaleh, S. M. Wasim, J. Galibert et al., Phil. Mag. B **65**, 843 (1992);

    G. Biskupski, Phil. Mag. B **65**, 723 (1992)

28. V. L. Nguyen, B. Z. Spivak, and B. I. Shklovskii, JETP Lett. **41**, 42 (1985)

29. B. L. Altshuler, A. G. Aronov, and D. E. Khmel'nitskii, JETP Lett. **36**, 195

30. W.L.Macmillan. *Phys. Rev. B.* **24**, 2739(1981)

31. A.I.Larkin and  D.E.Khmel'nitskii, *Sov. Phys. JETP* **56**, 647(1982)

32. H.F.Hess, K.de Conde, T.F.Rosenbaum and G.L.Salinger, *Phys. Rev. B.* **25**,

    5578(1982)

33. M.Reghu, Y.Cao, D.Moses and A.J.Heeger, *Phys. Rev. B* **47**, 1758(1993)

34. M.Reghu, C.O.Yoon, A.J.Heegar and Y.Cao, *Phys. Rev.B* **48**, 17685(1993)

35. M.Reghu, K.Vakiparta, C.O.oon, Y.Cao, D.Moses and A.J.Heegar, *Synth. Met*. **65**, 167 (1994)

36. D.E.Khmel'nitskii and A.I.Larkin, *Solid State Commun*. **39**, 1069(1981).

Table I. Comparison of boron doped samples with undoped one using X-ray diffraction technique.

| SAMPLE | PEAK($2\theta$) (Degree) | FWHM (Degree) | c (nm) | $x_B$ (%) |
|---|---|---|---|---|
| CB800M | 25.31 | 3.96 | 0.7028 | 25 |
| CB800M2 | 25.16 | 4.44 | 0.707 | 18 |
| CB800M4 | 25 | 5.41 | 0.7116 | 10 |
| C800 | 24.77 | 5.45 | 0.7178 | |
| CB900M | 24.2 | 5.87 | 0.7348 | |
| CB900M2 | 24.41 | 5.07 | 0.7284 | |
| CB900M4 | 24.89 | 4.11 | 0.714 | |
| C900 | 25 | 3.52 | 0.7116 | |
| CB700M4 | 24.9 | 5.73 | 0.713 | |
| C700 | 24.30 | 5.54 | 0.734 | |

Table II Comparison of theoretical constants for carbon samples with different doping level of boron and prepared at $900^0$C.

| Sample | CB900M | CB900M2 | CB900M4 | C900 |
|---|---|---|---|---|
| Type | Insulating | Insulating | Critical | Metallic |
| R(1.3)/R(300) | 37 | 15 | 2.66 | 1.16 |
| Resistivity at 300K ($\Omega$-cm) | 0.634 | 0.362 | 0.161 | 0.021 |
| $T_{Mott}$ (K) | 405 | 143 | | |
| $T_{ES}$ (K) | 21 | 11 | | |
| $T_{Mott}/T_{ES}$ | 19.3 | 13 | | |
| $T_{cross}$ (K) | 17K | 13K | | |
| $T_{CG}$ (K) | 0.506 | 0.179 | | |
| Coulomb Gap $\Delta_{CG}$ ( meV ) | 0.044 | 0.015 | | |
| Localization Length ($\alpha^{-1}$) ( nm) | 16.4 | 20.1 | | |
| $N(E_F)$ ($J^{-1}m^{-3}$) | $7.3 \times 10^{44}$ | $1.1 \times 10^{45}$ | | |
| $R_{hop, Mott} / \alpha^{-1}$ | $1.68 / T^{1/4}$ | $1.29 / T^{1/4}$ | | |
| $R_{hop, ES} / \alpha^{-1}$ | $1.14 / T^{1/2}$ | $0.82 / T^{1/2}$ | | |
| $\beta$ | | | 0.23 | |

Table III Comparison of theoretical constants for carbon samples with different doping level of boron and prepared at $800^{0}$C.

| Sample | CB800M | CB800M2 | CB800M4 | C800 |
|---|---|---|---|---|
| Type | Insulating | Critical | Critical | Critical |
| R(1.3)/R(300) | 42253 | 9.4 | 3.2 | 8 |
| Resistivity at 300K ($\Omega$-cm) | 1.93 | 0.5 | 0.3 | 0.48 |
| $T_{Mott}$ (K) | 1094 | | | |
| $T_{ES}$ (K) | 24 | | | |
| $T_{Mott}/T_{ES}$ | 19.3 | | | |
| $T_{cross}$ (K) | 17 | | | |
| $T_{CG}$ (K) | 1.36 | | | |
| Coulomb Gap $\Delta$CG ( meV ) | 0.12 | | | |
| Localization Length $\alpha^{-1}$ (nm) | 17.2 | | | |
| $N(E_F)$ ($J^{-1}m^{-3}$) | $2.34 \times 10^{44}$ | | | |
| $R_{hop, Mott} / \alpha^{-1}$ | $2.15 / T^{1/4}$ | | | |
| $R_{hop, ES} / \alpha^{-1}$ | $1.22 / T^{1/2}$ | | | |
| $\beta$ | -- | 0.22 | 0.60 | 0.95 |

Table IV Comparison of theoretical constants for carbon samples with different doping level of boron and prepared at $700^0$C. Samples having more boron (CB700M2 and CB700M) concentration were too insulating to carry out any conductivity measurements.

| Sample | CB700M4 | C700 |
|---|---|---|
| Type | Insulating | Insulating |
| R(4.2)/R(300) | 248270 | 194741 |
| Resistivity at 300K ($\Omega$-cm) | 18.5 | 13.3 |
| $T_{Mott}$ (K) | 8145 | 5470 |
| $T_{ES}$ (K) | 237 | 216 |
| $T_{Mott}/T_{ES}$ | 34.4 | 25.3 |
| $T_{cross}$ (K) | 54.4 | 53.2 |
| $T_{CG}$ (K) | 10.17 | 6.82 |
| $\Delta CG$ (meV) | 0.88 | 0.6 |
| $\alpha^{-1}$ (nm) | 1.1 | 1.9 |
| $N(E_F)$ ($J^{-1}m^{-3}$) | $1.2 \times 10^{47}$ | $3.46 \times 10^{46}$ |
| $R_{hop, Mott} / \alpha^{-1}$ | $3.56/ T^{1/4}$ | $3.2/ T^{1/4}$ |
| $R_{hop, ES} / \alpha^{-1}$ | $3.85 / T^{1/2}$ | $3.67 / T^{1/2}$ |

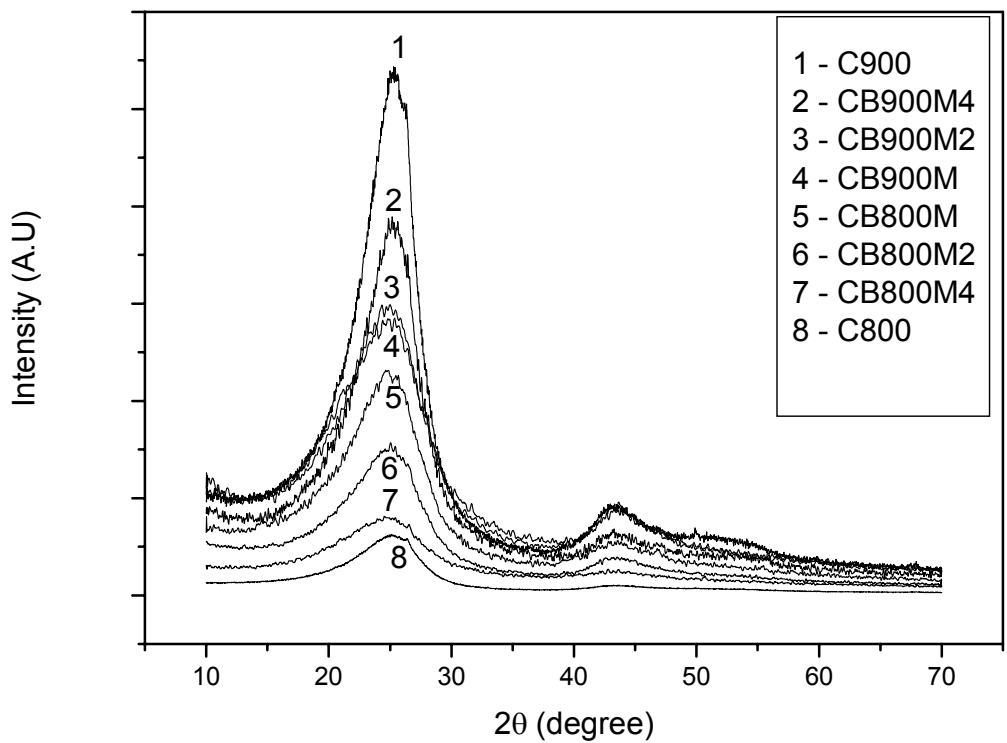

Figure 1. X-ray diffraction plot for a-conducting carbon film doped with boron of different doping level (Y-axis data is slightly displaced for the sake of clarity). The numbers corresponding to each x-ray diffraction graph corresponds to the sample labeled in the inset. The spectra of $700^0$C samples are similar to that of $800^0$C, so they are not shown here but the results are tabulated in table I.

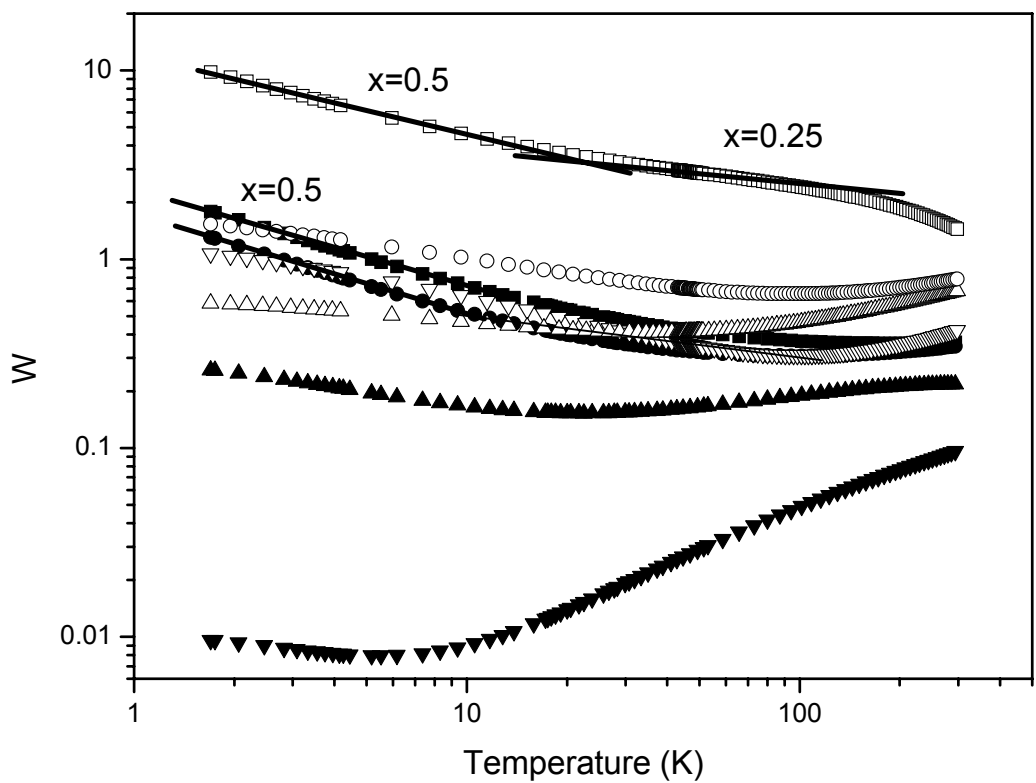

Figure 2.  Reduced activation energy plot for boron doped samples prepared at 900$^0$C

[(■)CB900M, (●)CB900M2, (▲)CB900M4, (▼)C900] and 800$^0$C [(□)CB800M,

(○)CB800M2, (△)CB800M4, (▽)C800]. The straight line fitting of the curve represents

ES VRH (x=0.5) and Mott VRH (x=0.25). Other curves are not fitted to these models because they lie in critical or metallic regime.

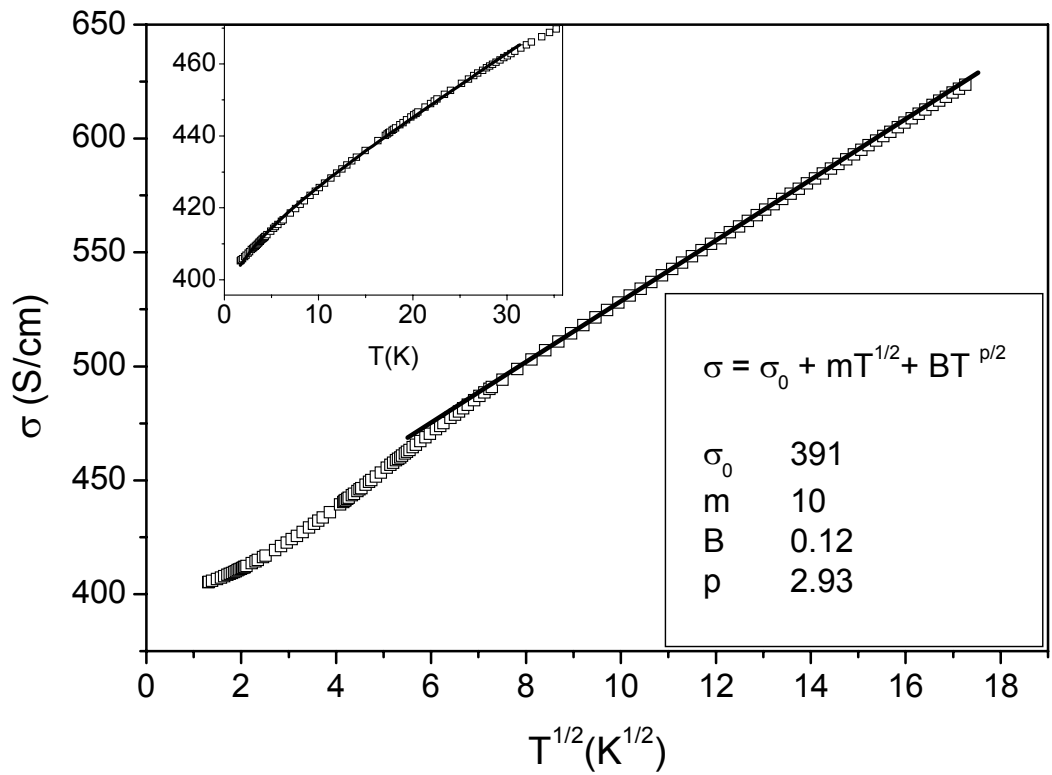

Figure 3. Temperature dependence of conductivity for C900 (metallic). The inset shows the $\sigma(T) = \sigma_0 + mT^{1/2} + BT^{p/2}$ variation below 30K. The values of constants are on the graph.

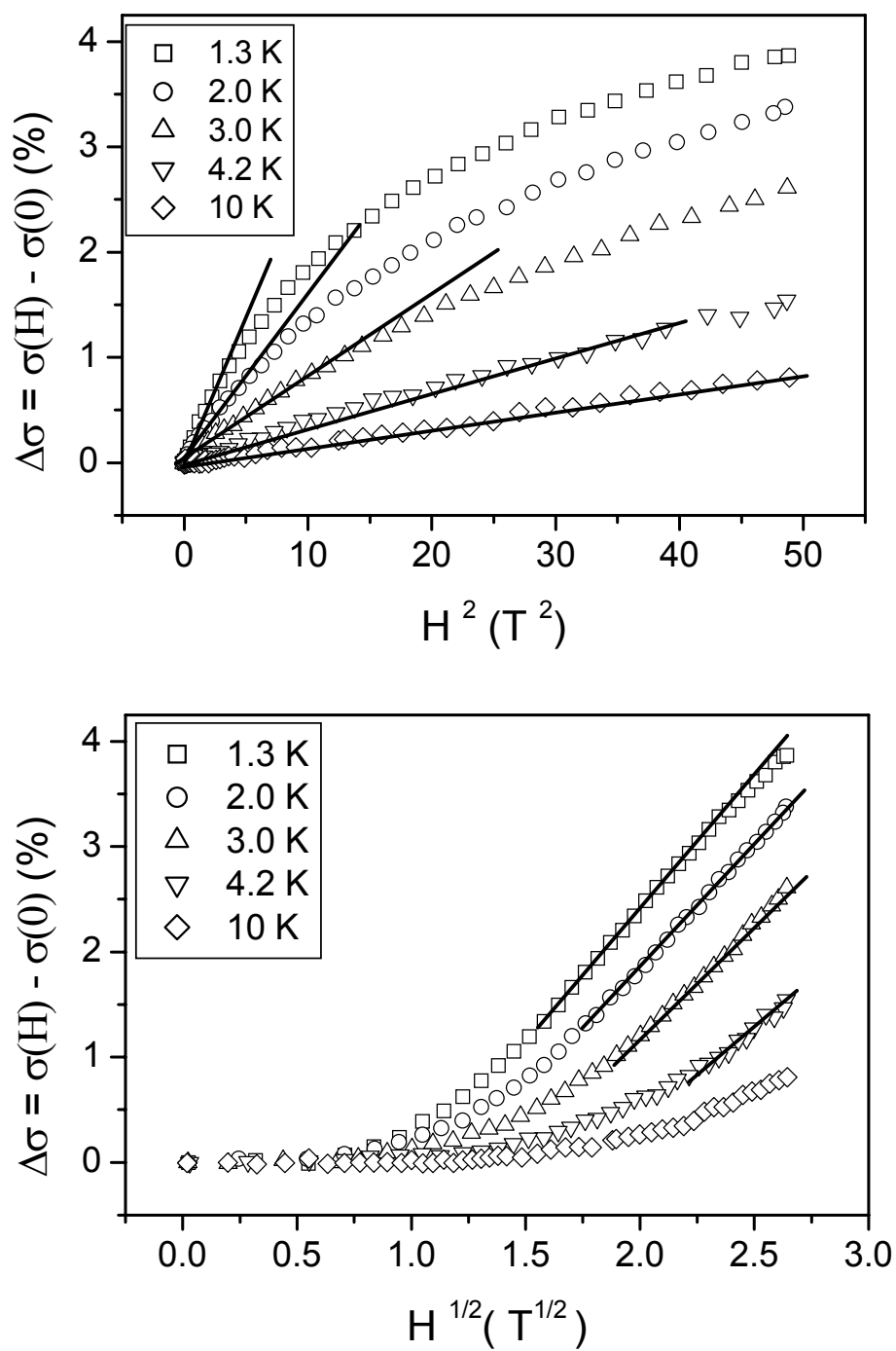

Figure 4. Behaviour of magnetoresistance for C900 at various temperatures. The linear

fitting represents $H^2$ dependence at low magnetic field (top) and $H^{1/2}$ dependence at high magnetic fields (below).

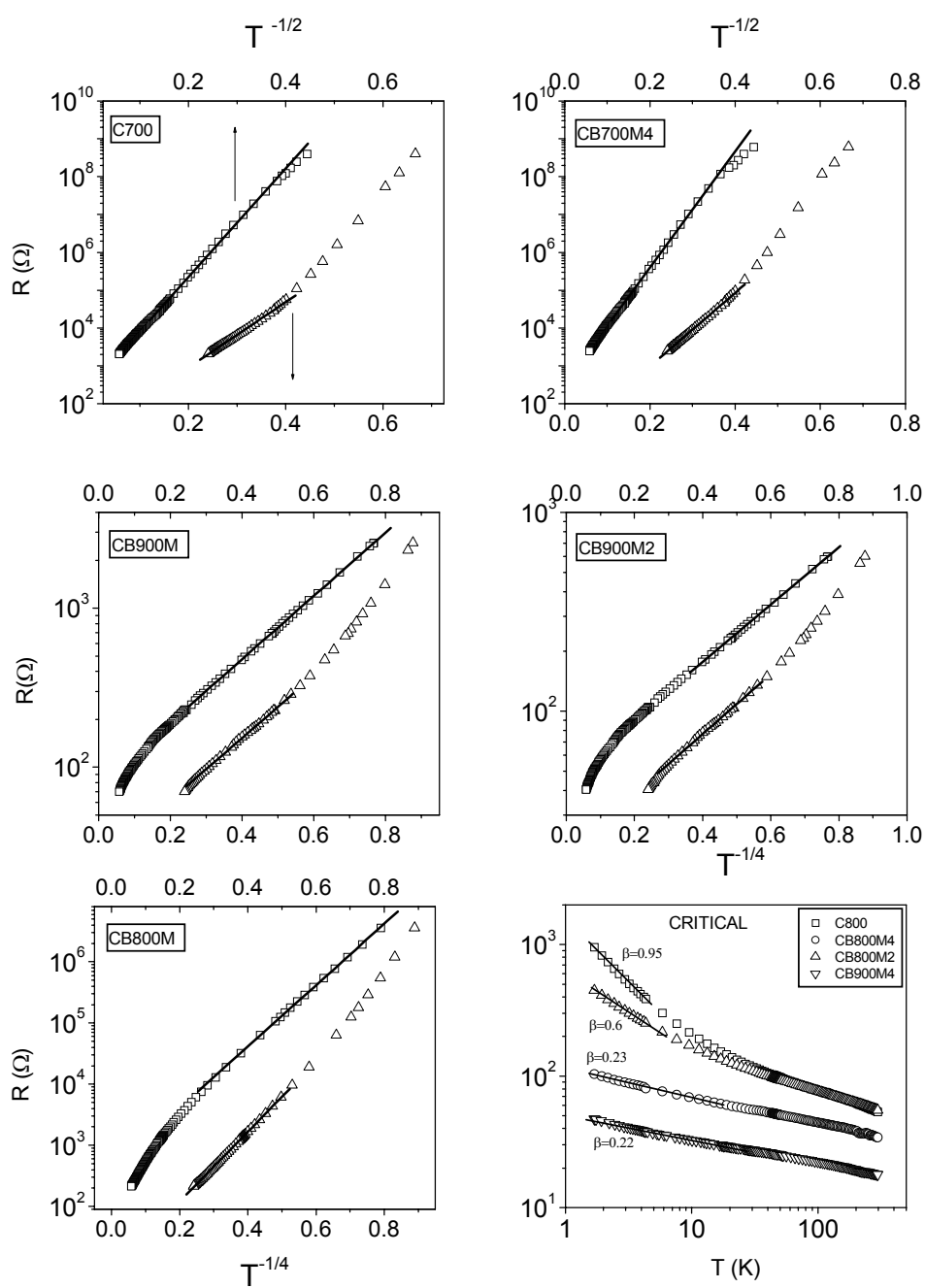

Figure. 5 Temperature dependence of resistivity for insulating samples CB900M, CB900M2, CB700M4, C700, CB800M and for critical samples C800, CB800M4, CB800M2 and CB900M4. In the graph, top x-axis is for $T^{-1/2}$ and the bottom x-axis is for $T^{-1/4}$ behaviour (except for critical samples).

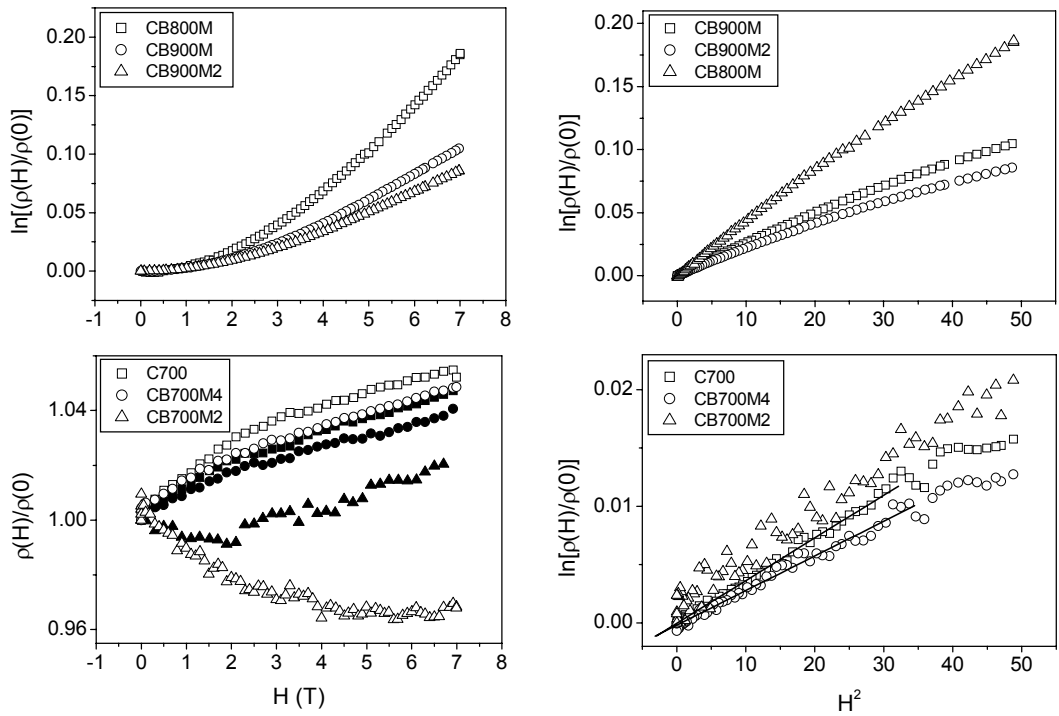

Figure 6. Square dependence of magnetoresisance at low magnetic field and linear dependence at high magnetic field for insulating samples CB800M, CB900M and CB900M2 at 4.2K (top). Also, shown is the variation of magnetoresistance for C700, CB700M4 and CB700M2 as a function of magnetic field. The magnetoresistance plot on the right is at 50K and on the left is at 15K (open symbols) and 20K (solid symbols).

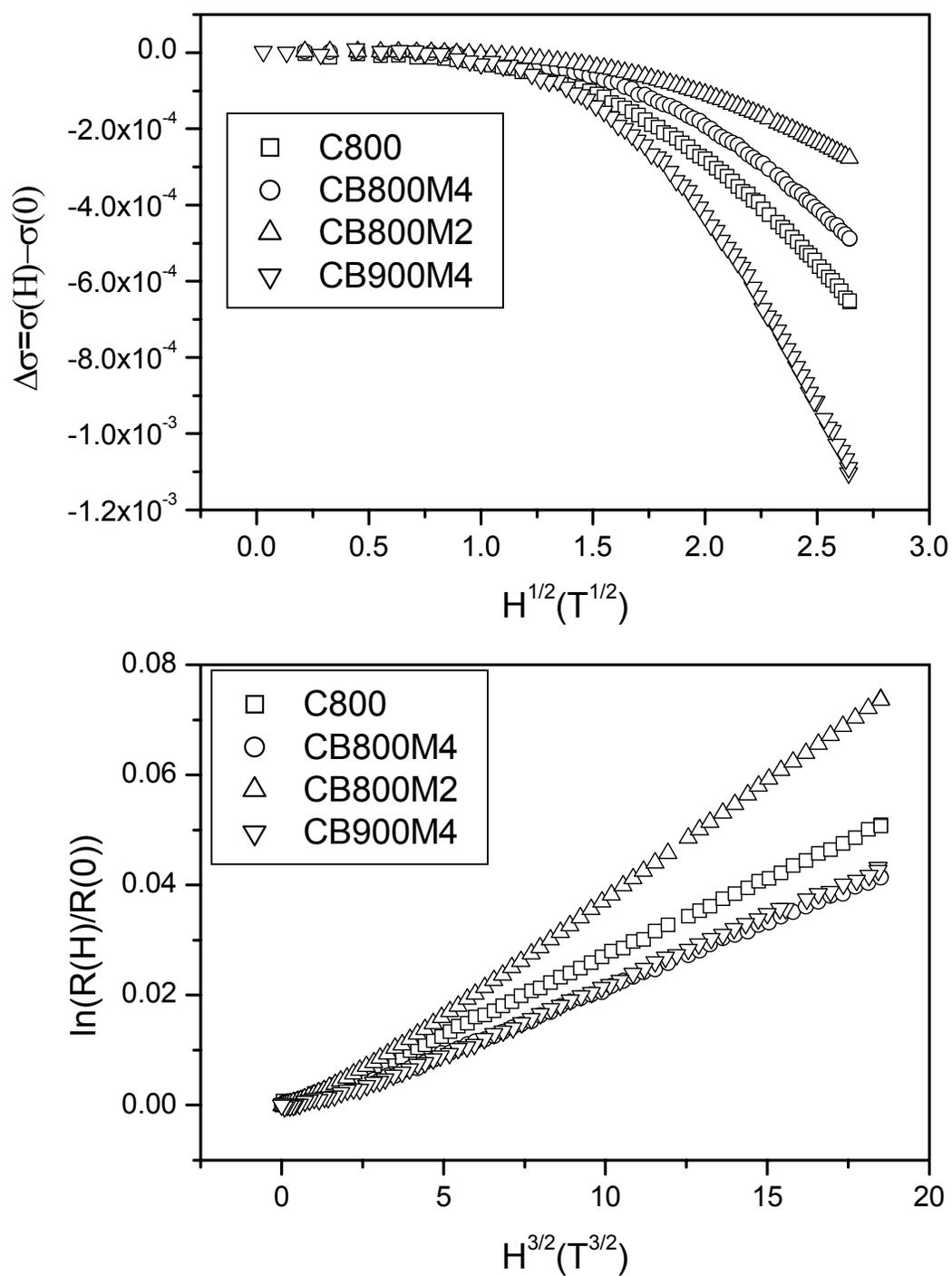

Figure 7.  Square root dependence of magnetoconductance (as predicted by theory) for critical samples at high magnetic field (top) and $H^{3/2}$ dependence (bottom) for the same.